\documentclass[twocolumn,american]{iopart}

\makeatletter
\usepackage{iopams}
\usepackage{setstack}

\makeatother

\usepackage{babel}

\makeatother

\usepackage{babel}

\begin{document}

\rapid{On 4-dimensional Lorentz-structures, Dark energy and Exotic smoothness }

\author{Torsten Asselmeyer-Maluga and Roland Mader}

\address{German Aerospace center, Rutherfordstr. 2, 12489 Berlin, Germany }

\ead{torsten.asselmeyer-maluga@dlr.de}

\author{Jerzy Kr\'ol}

\address{University of Silesia, Institute of Physics, ul. Uniwesytecka 4,
40-007 Katowice, Poland}

\ead{iriking@wp.pl}
\begin{abstract}
Usually, the topology of a 4-manifolds $M$ is restricted to admit
a global hyperbolic structure $\Sigma\times\mathbb{R}$. The result
was obtained by using two conditions: existence of a Lorentz structure
and causality (no time-like closed curves). In this paper we study
the influence of the smoothness structure to show its independence
of the two conditions. Then we obtain the possibility for a topology-change
of the 3-manifold $\Sigma$ keeping fix its homology. We will study
the example $S^{3}\times\mathbb{R}$ with an exotic differential structure
more carefully to show some implications for cosmology. Especially
we obtain an interpretation of the transition in topology as dark
energy.
\end{abstract}

\pacs{04.20.Gz, 98.80.Jk, 95.36.+x}

\submitto{\JPA}

\maketitle
A manifold admits a Lorentzian structure if (and only if) there is
a line element field (i.e. a non-vanishing vector field of arbitrary
sign, see Hawking and Ellis \cite{HawEll:94}). This condition is
equivalent to the existence of a codimension-1 foliation \cite{Law:74}.
Then a compact manifold admits a Lorentzian metric if the Euler characteristics
vanishes. In case of a compact 4-manifold one obtains a multiple-connected
manifold. But then one has the problem of causality (time loops etc.,
see also the topological censorship by Schleich and Witt \cite{FriedmanSchleichWitt93}).
Usually causality can be implemented on a non-compact 4-manifold.
Furthermore, detailed measurements of the background radiation by
the sattelites COBE and WMAP \cite{WMAPcompactSpace2008} enforces
us to assume that the space-like component is a compact 3-manifold
$\Sigma$. Both conditions, Lorentz structure and causality, can be
trivially fulfilled on 4-manifolds $\Sigma\times\mathbb{R}$. But,
is this choice unique?

\section{Lorentz structure and smoothness}

Non-compactness of the 4-manifold $\Sigma\times\mathbb{R}$ as the
result of the causality condition is a purely topological condition.
The second condition, the existence of a Lorentz structure, is related
to the structure of the tangent bundle $T(\Sigma\times\mathbb{R})$,
i.e. to the smoothness structure. 

The maximal differentiable atlas of a manifold is called its smoothness
structure. It is unique up to diffeomorphisms. In dimensions smaller
than 4 there is a unique smoothness structure \cite{Rad:25,Moi:52}
whereas in dimensions greater than 4 we have finitely many different
(=non-diffeomorphic) smoothness structures \cite{KirSie:77}. Only
in dimension 4 there is the possibility for infinite many different
smoothness structures with tremendous implication for quantum field
theory. For a deeper insight we refer to the book \cite{Asselmeyer2007}. 

If two manifolds are homeomorphic but non-diffeomorphic, they are
\textbf{exotic} to each other. The smoothness structure is called
an \textbf{exotic smoothness structure}. Among the smoothness structures
there is one distinguish element, the \textbf{standard structure}.
This structure is given by a smooth embedding of the 4-manifold into
some Euclidean space $\mathbb{R}^{N}$ for $N>7$. We remark further
that different smoothness structures have to represent different physical
situations leading to different measurable results. But it should
be stressed that \emph{exotic smoothness is not exotic physics.} Exotic
smoothness is a mathematical structure which should be further explored
to understand its physical relevance. 

In case of the Lorentz structure, we have to deal with the tangent
bundle $T(\Sigma\times\mathbb{R})$. We mentioned above that the existence
of a Lorentz structure is connected with the existence of a line element.
This line element exists only if the tangent bundle admits a splitting\[
T(\Sigma\times\mathbb{R})=\xi\oplus\chi\]
where $\chi$ is a 3-dimensional subbundle and $\xi$ is a 1-dimensional
subbundle of the tangent bundle. A section of the bundle $\xi$ is
the line element. Now let $\Sigma\times_{\Theta}\mathbb{R}$ be an
exotic 4-manifold, i.e. $\Sigma\times_{\Theta}\mathbb{R}$ is homeomorphic
to $\Sigma\times\mathbb{R}$ but not diffeomorphic to it. Usually
the cross product $\times$ will be only understand topologically.
Here we extend it to the smooth situation as well. But we know that
every 3-manifold $\Sigma$ has a unique smoothness structure. Therefore
$\Sigma\times\mathbb{R}$ represents smoothly the standard structure
and we choose $\Sigma\times_{\Theta}\mathbb{R}$ to indicate the exotic
smoothness structure. From this point of view we have a big difference
between the bundle $T(\Sigma\times\mathbb{R})$ and $T(\Sigma\times_{\Theta}\mathbb{R})$.
The first bundle admits a splitting\[
T(\Sigma\times\mathbb{R})=T\Sigma\times T\mathbb{R}\]
whereas the second is not splittable in that manner. But it is known
that the tangent bundle of every 3-manifold is trivial \cite{MSt:74},
i.e. $T\Sigma=\Sigma\times\mathbb{R}^{3}$. From this fact we obtain
a splitting \[
T(\Sigma\times_{\Theta}\mathbb{R})=\xi\oplus\chi\]
as needed for a Lorentz structure. This splitting is induced by the
embedding $\Sigma\to\Sigma\times_{\Theta}\mathbb{R}$ and by the bundle
splitting\[
T\Sigma=\Sigma\times\mathbb{R}^{3}=(\Sigma\times\mathbb{R}^{2})\oplus(\Sigma\times\mathbb{R}^{1})\]
into a 1- and 2-dimensional subbundle. Therefore the exotic 4-manifold
$\Sigma\times_{\Theta}\mathbb{R}$ admits also a Lorentz structure
but of a different kind. Now one has to study a codimension-1 foliation
of the 3-manifold $\Sigma$, i.e. one has the usual splitting (3-space$\times$time)
for submanifolds (the leaves) of $\Sigma$ only.

\section{The Example $S^{3}\times\mathbb{R}$\label{sec:The-Example}}

In this section we will study a specific example. For simplicity we
choose $S^{3}\times\mathbb{R}$ with a global time function so that
$S^{3}\times\left\{ t\right\} $ is the leaf for every $t\in\mathbb{R}$
in this section. So, we obtain a foliation of $S^{3}\times\mathbb{R}$.
Now we discuss the change of the smoothness structure leading to a
change of the foliation for $S^{3}\times\mathbb{R}$. One of the first
examples of an exotic smoothness structure on this manifold was given
by Freedman \cite{Fre:79} using the Poincare sphere. Lets denote
the exotic $S^{3}\times\mathbb{R}$ by $S^{3}\times_{\Theta}\mathbb{R}$.
The foliation of $S^{3}\times_{\Theta}\mathbb{R}$ contains a Poincare
sphere as smooth cross section (see Theorem 4 in \cite{Fre:79}).
In \ref{sec:Constructing-exotic-S3xR}, we will give the details of
the construction. Here we will only present a short outline. One starts
with a homology 3-sphere $P$, i.e. a compact 3-manifold $P$ with
the same homology as the 3-sphere but non-trivial fundamental group,
see \ref{sub:Homology-3-Spheres}. The Poincare sphere is one example
of a homology 3-manifold. Now we consider the 4-manifold $P\times[0,1]$
with the same fundamental group $\pi_{1}(P\times[0,1])=\pi_{1}(P)$.
By a special procedure (the plus construction see \cite{Mil:61,Ros:94}),
one can ''kill'' the fundamental group $\pi_{1}(P)$ in the intorior
of $P\times[0,1]$. This procedure will result in a 4-manifold $W$
with boundary $\partial W=-P\sqcup S^{3}$ ($-P$ with opposite orientation),
a so-called cobordism between $P$ and $S^{3}$. The gluing $-W\cup_{P}W$
along $P$ with the boundary $\partial(-W\cup_{P}W)=-S^{3}\sqcup S^{3}$
defines one piece of the exotic $S^{3}\times_{\Theta}\mathbb{R}$.
The whole construction can be extended to both directions to get the
desired exotic $S^{3}\times_{\Theta}\mathbb{R}$ (see the \ref{sec:Constructing-exotic-S3xR}
for details). There is one critical point in the construction: the
4-manifold $W$ is not a smooth manifold. As Freedman \cite{Fre:82}
showed, the 4-manifold $W$ always exists topologically but by a result
of Gompf \cite{Gom:89} (using Donaldson \cite{Don:83}) not smoothly.
The Poincare sphere or the Brieskown sphere $\Sigma(2,3,7)$ are examples
of homology 3-spheres $P$ leading to a non-smooth $W$ whereas the
Brieskorn sphere $\Sigma(2,5,7)$ produces a smooth 4-manifold $W$. 

The 4-manifold $-W\cup_{P}W$ is also non-smoothable and we will get
a smoothness structure only for the non-compact $S^{3}\times_{\Theta}\mathbb{R}$
(see \cite{Qui:82}). But $S^{3}\times_{\Theta}\mathbb{R}$ contains
$-W\cup_{P}W$ with the smooth cross section $P$. From the physical
point of view we interpret $-W\cup_{P}W$ as a time line of a cosmos
starting as 3-sphere changing to the homology 3-sphere $P$ and changing
back to the 3-sphere. But this process is part of every exotic smoothness
structure $S^{3}\times_{\Theta}\mathbb{R}$, i.e. we obtain\\
\emph{In the exotic $S^{3}\times_{\Theta}\mathbb{R}$ we have
a change of the spatial topology from the 3-sphere to some homology
3-sphere.}\\
But this conclusion is only part of the story. In cosmology one
usually consider an isotropic and homogenous model for the cosmos.
Then the spatial 3-manifold has to admit a homogenous geometry with
constant curvature. Obviously the 3-sphere has positive curvature
as well as the Poincare 3-sphere (using Thurstons geometrization conjecture
as proven by Perelman). So, from the geometrical point of view nothing
changes for an exotic $S^{3}\times_{\Theta}\mathbb{R}$ constructed
from the Poincare sphere. But consider now the homology sphere $\Sigma(8_{10})$
with hyperbolic geometry constructed in the \ref{sec:Constructing-hyperbolic-hom-S3}.
Then the corresponding $S^{3}\times_{\Theta}\mathbb{R}$ contains
a change of the topology and geometry from spherical 3-sphere to a
hyperbolic $\Sigma(8_{10})$ and back. \\
\emph{In the exotic $S^{3}\times_{\Theta}\mathbb{R}$ there is
a change of the geometry from the spherical 3-sphere to some homology
3-sphere with geometry of postive or negative curvature.}

\section{Exotic cosmology and Dark energy}

Our choice of the example in the previous subsection was not arbitrary.
Given a compact 3-manifold $\Sigma$, the connected sum%
\footnote{Let $M,N$ be a compact $n$-manifolds. The connected sum $M\#N$
is the procedure to cut out a disk $D^{n}$ from the interior $int(M)\setminus D^{n}$
and $int(N)\setminus D^{n}$ with the boundaries $S^{n-1}\sqcup\partial M$
and $S^{n-1}\sqcup\partial N$, respectively, and glue them together
(by a smooth map) along the common boundary component $S^{n-1}$.
The boundary $\partial(M\#N)=\partial M\sqcup\partial N$ is the disjoint
union of the boundaries $\partial M,\partial N$. %
} $\Sigma\#S^{3}$ is diffeomorphic to $\Sigma$. Therefore $\Sigma\times\mathbb{R}$
contains in some sense the example $S^{3}\times\mathbb{R}$. Especially
we can generalize the conclusions above to this case as well. Now
we will study the geometry and topology changing process more carefully.
Lets consider the Robertson-Walker metric (with $c=1$)\[
ds^{2}=dt^{2}-a(t)^{2}h_{ik}dx^{i}dx^{k}\]
with the scaling function $a(t)$. At first we assume a spacetime
$S^{3}\times\mathbb{R}$ with increasing function $a(t)$ fulfilling
the Friedman equations\begin{eqnarray}
\left(\frac{\dot{a}(t)}{a(t)}\right)^{2}+\frac{k}{a(t)^{2}} & = & \kappa\frac{\rho}{3}\label{eq:friedman-1}\\
2\left(\frac{\ddot{a}(t)}{a(t)}\right)+\left(\frac{\dot{a}(t)}{a(t)}\right)^{2}+\frac{k}{a(t)^{2}} & = & -\kappa p\label{eq:friedman-2}\end{eqnarray}
derived from Einsteins equation\begin{equation}
R_{\mu\nu}-\frac{1}{2}g_{\mu\nu}R=\kappa T_{\mu\nu}\label{eq:Einstein-equation}\end{equation}
with the gravitational constant $\kappa$ and the energy-momentum
tensor of a perfect fluid\begin{equation}
T_{\mu\nu}=(\rho+p)u_{\mu}u_{\nu}-pg_{\mu\nu}\label{eq:perfect-fluid-EM-tensor}\end{equation}
with the (time-dependent) energy density $\rho$ and the (time-dependent)
pressure $p$. The spatial cosmos has the scalar curvature $^{3}R$\[
^{3}R=\frac{k}{a^{2}}\]
from the metric $h_{ik}$ and we obtain the 4-dimensional scalar curvature
$R$\begin{equation}
R=\frac{6}{a^{2}}\left(\ddot{a}\cdot a+\dot{a}^{2}+k\right)\label{eq:4-dim-scalar-curvature}\end{equation}
or\[
R=3\kappa(\rho-3p)\]
with the acceleration\[
\frac{\ddot{a}}{a}=-\frac{\kappa}{6}\left(\rho+3p\right)\,.\]
Let us consider the model $S^{3}\times\mathbb{R}$ with positive spatial
curvature $k=+1$. In case of dust matter ($p=0$) only, one obtains
a closed universe. But for negative spatial curvature $k=-1$, one
has on open universe. Now we consider our model of an exotic $S^{3}\times_{\Theta}\mathbb{R}$,
see section \ref{sec:The-Example}. As explained above, the foliation
of $S^{3}\times_{\Theta}\mathbb{R}$ must contain a homology 3-sphere
$\Sigma(8_{10})$ with negative curvature. But then we have a transition
from a space with positive curvature to a space with negative curvature
and back. In the \ref{sec:Constructing-exotic-S3xR}, this period
was denoted by $-W_{1}\cup W_{1}$ where both $W_{1}$ were glued
along $\Sigma(8_{10})$. At first, in all ''later'' versions of the
spatial space we have the space $\Sigma(8_{10})$. Secondly, hyperbolic
3-manifolds like $\Sigma(8_{10})$ have a special property, called
Mostow rigidity \cite{Mos:68}. It means, that a hyperbolic 3-manifold
can not be scaled, i.e. the volume or the curvature is a topological
invariant. Therefore, if the hyperbolic 3-manifold $\Sigma$ is part
of the space then this part has a constant curvature. We will use
this property in the following.

Let us assume that the period of the transition starts at time $t_{0}$
with negligible length. So, for a time $t>t_{0}$ we have the effective
equation\[
2\left(\frac{\ddot{a}(t)}{a(t)}\right)+\left(\frac{\dot{a}(t)}{a(t)}\right)^{2}+\frac{1}{a(t)^{2}}-\frac{1}{(a(t_{0}))^{2}}=-\kappa p\]
with the constant $\Lambda=(a(t_{0}))^{-2}$ proportional to the constant
curvature of $\Sigma$. The first equation (\ref{eq:friedman-1})
is modified in the same manner. The volume of $\Sigma(8_{10})$ is
proportional to $(a(t_{0}))^{3}$. Finally we obtain the acceleration\[
\frac{\ddot{a}}{a}=\frac{\Lambda}{2}-\frac{\kappa}{6}\left(\rho+3p\right)\,.\]
This acceleration can be positive, i.e. for a time $t\gg t_{0}$ the
term $\Lambda$ dominates and we obatin an accelerated expansion.

\section{Conclusion}

In this paper we discuss spacetime models of the universe with an
exotic smoothness structure. In four dimensions there is an infinity
of possible exotic structures. So, it will be rather a surprise if
our universe admits the standard smoothness structure. Here we paid
special attention to the model $S^{3}\times\mathbb{R}$ and modify
it to the exotic $S^{3}\times_{\Theta}\mathbb{R}$. Then we obtain
a topological transition from the space $S^{3}$ to a hyperbolic homology
3-sphere $\Sigma(8_{10})$ (see \ref{sec:Constructing-hyperbolic-hom-S3}).
This transition can be interpreted as dark energy which forces the
expansion to accelerate. Therefore exotic smoothness can serve as
another way to obtain the accelerated expansion. Of course the value
of the cosmological constant $\Lambda$ is not determined yet. But
this value must be much smaller then every estimate of quantum field
theory. In our forthcoming work we will try explain the value.

\appendix

\section{Homology 3-spheres\label{sub:Homology-3-Spheres}}

A homology 3-sphere $\Sigma$ is a compact, connected 3-manifold without
boundary having the same homology as the 3-sphere $S^{3}$. This definition
implies the homology groups of $\Sigma$: \[
H_{0}(\Sigma)=H_{3}(\Sigma)=\mathbb{Z},\quad H_{1}(\Sigma)=H_{2}(\Sigma)=0\]
This definition seem to restrict the 3-manifolds very strong but there
is one invariant characterizing 3-manifolds rather uniquely: the fundamental
group $\pi_{1}(\Sigma)$ (closed curves up to homotopy). The first
example of a homology 3-sphere was constructed by Poincar\'e using
the binary icosaeder group $I^{*}=\langle s,t\:|\: s^{5}=(st)^{2},t^{3}=(st)^{2}\rangle$,
i.e. the group of sequences generated by $s,\, t$ and its inverses
$s^{-1},\, t^{-1}$ restricted by the relations $s^{5}=(st)^{2},\, t^{3}=(st)^{2}$.
The group action $S^{3}\times I^{*}\to S^{3}$ is free and one has
the equivalence classes $S^{3}/I^{*}$ forming a smooth manifold,
the \emph{Poincar\'e sphere}. The fundamental group is given by $\pi_{1}(S^{3}/I^{*})=I^{*}$.
The group $I^{*}$ has a very important property: $I^{*}$ is perfect.
For tht purpose we define the commutator $[s,\, t]=sts^{-1}t^{-1}$
in a group $G$ and denote with $[G,\, G]$ the subgroup generated
by the commutators of $G$. One calls a group \emph{perfect} iff $G=[G,\, G]$.
Now we need a relation between the first homology group and the fundamental
group given by \[
H_{1}(M)=\frac{\pi_{1}(M)}{[\pi_{1}(M),\pi_{1}(M)]}\]
Then the vanishing of $H_{1}(M)$ is equivalent to a perfect fundamental
group $\pi_{1}(M)=[\pi_{1}(M),\pi_{1}(M)]$. Therefore we obtain from
the compactness $H_{0}(S^{3}/I^{*})=H_{3}(S^{3}/I^{*})=\mathbb{Z}$
and $H_{1}(S^{3}/I^{*})=H_{2}(S^{3}/I^{*})=0$ from $I^{*}=[I^{*},\, I^{*}]$
and duality. The group $I^{*}$ has 120 elements and is the \emph{only
finite, perfect} group.

\section{Constructing exotic $S^{3}\times\mathbb{R}$'s\label{sec:Constructing-exotic-S3xR}}

Given a homology 3-sphere $\Sigma$ which do not bound a contractable,
smooth 4-manifold. According to Freedman \cite{Fre:82}, every homology
3-sphere bounds a contractable, topological 4-manifold but not every
of these 4-manifolds is smoothable. Now we consider the following
pieces: $W_{1}$ as cobordism between $\Sigma$ and its one-point
complement $\Sigma\setminus pt.$ as well $W_{2}$ as cobordism between
$\Sigma\setminus pt.$ and $\Sigma\setminus pt.$. The manifold $W=\ldots\cup-W_{2}\cup-W_{2}\cup-W_{1}\cup W_{1}\cup W_{2}\cup W_{2}\cup\ldots$
(see \cite{Fre:79}) is homeomorphic to $S^{3}\times\mathbb{R}$ (using
the proper h-cobordism theorem in \cite{Fre:82}) but not diffeomorphic
to it, i.e. $W=S^{3}\times_{\Theta}\mathbb{R}$. The construction
of the pieces $W_{1},W_{2}$ rely heavily on the concept of a Casson
handle. For that purpose we consider a Casson handle $CH$ and its
3-stage tower $T_{3}^{1}$. By the embedding theorem of Freedman (see
\cite{Fre:79}, Theorem 1), one can construct another 3-stage tower
$T_{3}^{1}$ inside of $T_{3}^{0}$ (increase the number of self-intersections
of the core). This process can be done infinitely. Lets take an example
of an homology 3-sphere $\Sigma$ constructed in the next section.
The fundamental group is generated by one generator, i.e. we need
a single Casson handle only to kill this generator by defining an
embedding $T_{3}^{0}\hookrightarrow\Sigma\times[0,1]$. Now by killing
an arc in each 3-stage tower $T_{3,arc}^{0}=T_{3}^{0}\setminus\left\{ \mbox{arc}\right\} $
and by killing a line $T_{3,line}^{0}=T_{3}^{0}\setminus\left\{ \mbox{line}\right\} $we
can form the desired cobordisms $W_{1},W_{2}$ above: $W_{1}=\Sigma\times[0,1]\setminus\bigcap_{i=0}^{\infty}T_{3,arc}^{i}$
and $W_{2}=\Sigma\times[0,1]\setminus\bigcap_{i=0}^{\infty}T_{3,line}^{i}$
completing the construction of the exotic $S^{3}\times\mathbb{R}$.
Furthermore we obtain the smooth cross section $\Sigma$ in the part
$-W_{1}\cup W_{1}$ of $W$.

\section{Constructing a hyperbolic homology 3-sphere\label{sec:Constructing-hyperbolic-hom-S3}}

Here we will construct one example of a hyperbolic homology 3-sphere
which does not bound a contractable 4-manifold. Then we can use the
procedure above to get an exotic $S^{3}\times_{\Theta}\mathbb{R}$.
For that purpose we consider a knot $K$, i.e. a smooth embedding
$S^{1}\to S^{3}$. This knot $K$ can be thicken to $N(K)=K\times D^{2}$
and one obtains the knot complement $C(K)=S^{3}\setminus N(K)$ with
boundary $\partial C(K)=T^{2}$. By the attachment of a solid torus
$D^{2}\times S^{1}$ using a $-1$ Dehn twist, one obtains the manifold
$\Sigma=C(K)\cup(D^{2}\times S^{1})$, a homology 3-sphere \cite{Rol:76}.
The geometry of $\Sigma$ is determined by the knot complement $C(K)$.
A fundamental result of Thurston \cite{Thu:97} states that most knot
complements are hyperbolic 3-manifolds. Examples are the figure-8
knot $4_{1}$, the 3-twist knot $5_{2}$ or the knot $8_{10}$ (in
Rolfsen notation \cite{Rol:76}). Therefore we have to look for a
knot inducing a hyperbolic knot complement and leading to a homology
3-sphere which does not bound a smooth contractable 4-manifold. As
Freedman \cite{Fre:82} showed, every homology 3-sphere bounnd a contractable,
topological 4-manifold. But Donaldson \cite{Don:83} found the first
example, the Poincare sphere, of a homology 3-sphere which fails to
do it. The Poincare sphere is generated by the trefoil knot $3_{1}$
using the procedure above. Every homology 3-sphere homology-cobordant
to the Poincare sphere has the same property \cite{Gordon1975}. From
the knot-theoretical point of view, we have to look for a knot concordant
to the trefoil knot \cite{Livingston1981}. One example is given by
the knot $8_{10}$ (see \cite{Livingston2004}). Then the homology
3-sphere $\Sigma$ constructed from this knot has all desired properties.


\begin{thebibliography}{10}

\bibitem{HawEll:94}
S.W Hawking and G.F.R. Ellis.
\newblock {\em The Large Scale Structure of Space-Time}.
\newblock Cambridge University Press, 1994.

\bibitem{Law:74}
H.B. Lawson.
\newblock Foliations.
\newblock {\em BAMS}, 80:369 -- 418, 1974.

\bibitem{FriedmanSchleichWitt93}
J.L. Friedman, K.~Schleich, and D.M. Witt.
\newblock Topological censorship.
\newblock {\em Phys. Rev. Lett.}, 71:1486--1489, 1993.
\newblock Erratum-ibid. 75 (1995) 1872; arXiv:gr-qc/9305017.

\bibitem{WMAPcompactSpace2008}
B.F. Roukema, Z.~Bulinski, A.~Szaniewska, and N.E. Gaudin.
\newblock The optimal phase of the generalised poincare dodecahedral space
  hypothesis implied by the spatial cross-correlation function of the {WMAP}
  sky maps.
\newblock {\em Astron. Astrophysics}, {\bf 486}:55--74, 2008.
\newblock arXiv:0801.0006 [astro-ph].

\bibitem{Rad:25}
T.~Rado.
\newblock {\"U}ber den {B}egriff der {R}iemannschen {F}l{\"a}che.
\newblock {\em Acta Litt. Scient. Univ. Szegd}, {\bf 2}:101--121, 1925.

\bibitem{Moi:52}
E.~Moise.
\newblock Affine structures on 3-manifolds.
\newblock {\em Ann. Math.}, {\bf 56}:96--114, 1952.

\bibitem{KirSie:77}
R.~Kirby and L.C. Siebenmann.
\newblock {\em Foundational essays on topological manifolds, smoothings, and
  triangulations}.
\newblock Ann. Math. Studies. Princeton University Press, Princeton, 1977.

\bibitem{Asselmeyer2007}
T.~Asselmeyer-Maluga and C.H. Brans.
\newblock {\em Exotic {S}moothness and {P}hysics}.
\newblock WorldScientific Publ., Singapore, 2007.

\bibitem{MSt:74}
J.~Milnor and J.~Stasheff.
\newblock {\em Characteristic Classes}.
\newblock Ann. Math. Studies,76. Princeton Univ. Press, Princeton, N.J., 1974.

\bibitem{Fre:79}
M.H. Freedman.
\newblock A fake {$S^3\times R$}.
\newblock {\em Ann. of Math.}, {\bf 110}:177--201, 1979.

\bibitem{Mil:61}
J.~Milnor.
\newblock A procedure for killing the homotopy groups of differentiable
  manifolds.
\newblock In {\em Proc. Symp. in Pure Math. 3 (Differential Geometry)}, pages
  39--55. Amer. Math. Soc., 1961.

\bibitem{Ros:94}
J.~Rosenberg.
\newblock {\em Algebraic {K}-theory and its application}.
\newblock Springer, 1994.

\bibitem{Fre:82}
M.H. Freedman.
\newblock The topology of four-dimensional manifolds.
\newblock {\em J. Diff. Geom.}, {\bf 17}:357 -- 454, 1982.

\bibitem{Gom:89}
R.~Gompf.
\newblock Periodic ends and knot concordance.
\newblock {\em Top. Appl.}, {\bf 32}:141--148, 1989.

\bibitem{Don:83}
S.~Donaldson.
\newblock An application of gauge theory to the topology of 4-manifolds.
\newblock {\em J. Diff. Geom.}, {\bf 18}:269--316, 1983.

\bibitem{Qui:82}
F.~Quinn.
\newblock Ends of {M}aps {III}: dimensions 4 and 5.
\newblock {\em J. Diff. Geom.}, {\bf 17}:503 -- 521, 1982.

\bibitem{Mos:68}
G.D. Mostow.
\newblock Quasi-conformal mappings in $n$-space and the rigidity of hyperbolic
  space forms.
\newblock {\em Publ. Math. IHÉS}, {\bf 34}:53--104, 1968.

\bibitem{Rol:76}
D.~Rolfson.
\newblock {\em Knots and Links}.
\newblock Publish or Prish, Berkeley, 1976.

\bibitem{Thu:97}
W.~Thurston.
\newblock {\em Three-Dimensional Geometry and Topology}.
\newblock Princeton University Press, Princeton, first edition, 1997.

\bibitem{Gordon1975}
M.A. Gordan.
\newblock Knots, homology spheres, and contractible 4-manifolds.
\newblock {\em Topology}, {\bf 14}:151--172, 1975.

\bibitem{Livingston1981}
Ch. Livingston.
\newblock Homology cobordisms of 3-manifolds, knot concordances, and prime
  knots.
\newblock {\em Pacific J. Math.}, {\bf 94}:193--206, 1981.

\bibitem{Livingston2004}
Ch. Livingston.
\newblock The concordance genus of knots.
\newblock {\em Alg. Geom. Top.}, {\bf 4}:1--22, 2004.

\end{thebibliography}
\end{document}